\documentclass[12pt]{book}

\usepackage[dvips]{graphicx,color}
\usepackage{makeidx,tsukuba}

\makeauthorindex
\makeindex

\begin{document}

\BookTitle{\itshape The 28th International Cosmic Ray Conference}
\CopyRight{\copyright 2003 by Universal Academy Press, Inc.}
\pagenumbering{arabic}

\chapter{
Intrinsic Spectra of the TeV Blazars Mrk 421 and Mrk 501}

\author{%
%
%
F.~Krennrich$^1$ \& E.~Dwek$^2$ \\
{\it (1) Department of Physics and Astronomy, Iowa State University
               Ames, IA, 50011, USA}
{\it (2) NASA Goddard Space Flight Center, Greenbelt, MD 20771 USA}
}

\section*{Abstract}
Energy spectra of $\rm \gamma$-ray blazars may contain an imprint from the
cosmic infrared background radiation due to $\rm \gamma$-ray absorption
(pair-production) by soft photons constituting the extragalactic background light (EBL). 
The signature of this imprint depends on the spectral shape of the EBL.
In this work we correct the observed  spectra of Mrk~421 and  Mrk~501 for 
absorption using different possible realizations of the EBL, consistent with the 
most recent detections and limits. 
We present the intrinsic $\gamma-$ray spectrum of these sources for the 
different EBL scenarios. These spectra reveal their true peak energy and 
luminosities, which provide important information on the nature and physical 
characteristics of the particle acceleration mechanism operating in  these sources. 
 
\section{Introduction}
Observations of blazars  at TeV $\rm \gamma$-ray energies are useful to probing
non-thermal phenomena in  jets of active galactic nuclei (AGNs) utilizing 
the highest energy  photons currently available to astronomy.  
Very high energy $\rm \gamma$-rays traveling cosmological distances are  attenuated
by  the diffuse extragalactic background light (EBL) via pairproduction (Gould \& Schr\' eder 
1967; Stecker et al. 1992).    
Hence, the discussion of TeV blazar spectra in the context of emission models and
multiwavelength studies requires a correction  for  $\rm \gamma$-ray absorption effects.   

In addition, TeV $\rm \gamma$-ray studies also provide strong constraints to the EBL
density  in  the wavelength regime  of 0.1~$\rm \mu$m - 30~$\rm \mu$m. 
The EBL is part of the extragalactic background radiation that ranges from $\rm 10^{-7}$~eV 
(radio background) to almost   $\rm 10^{11}$~eV.  Although the diffuse radiation is 
dominated by the cosmic microwave background,   the EBL   is the second most dominant 
form of  electromagnetic energy throughout the universe.    The EBL originates  from the time of 
structure/galaxy formation and evolution  (Partridge \& Peebles 1967;   Primack 1999)
and contains potentially important cosmological information about how galaxies formed.
Direct measurements of the EBL in the mid infrared are extremely difficult due to the 
presence of strong foreground zodiacal emission consisting of scattered light and 
thermal emission from interplanetary dust particles (Kelsall et al. 1998). 
For an extensive review on this,  see Hauser \& Dwek (2001).

In this paper we focus on the extraction of the intrinsic energy spectra of the two
most prominent TeV blazars, Mrk~421 and  Mrk~501  which have been measured with good
statistical precision at energies  between 260~GeV -- 20~TeV (Aharonian et a. 1999, 2002; Samuelson et al.
1998; Krennrich et al. 2001).    The data of the  relevant diffuse radiation 
fields (0.1~$\rm \mu$m - 100~$\rm \mu$m) consist largely of  upper and lower limits and few 
3$\rm \sigma$  level detections (Hauser \& Dwek 2001; Dwek \& Krennrich 2003).  In fact, at the mid-infrared
wavelenghts the   diffuse photon  densities are uncertain  within a factor of a few, resulting 
in significant  uncertainties in estimates of absolute luminosity of Mrk~421  and Mrk~501.  
However, the shape of the TeV spectra of nearby blazars below 10~TeV is less susceptible to the 
wide  range of possible EBL scenarios,  since all scenarios indicate a drop in 
the photon density between 1~$\rm \mu$m - 10~$\rm \mu$m, the most effective target  for 
absorption of  1~TeV  - 10~TeV $\rm \gamma$-rays.    Hence,  meaningful
peak energy measurements of the intrinsic TeV spectra are possible for nearby blazars
such as Mrk~501 and Mrk~421.
\begin{figure}[t]
  \begin{center}
    \includegraphics[height=12pc]{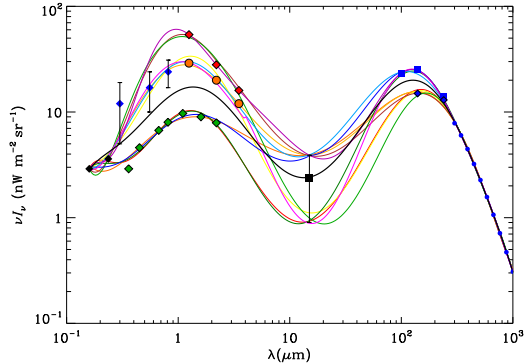}
  \end{center}
  \vspace{-0.5pc}
  \caption{A range of extragalactic background light scenarios are shown that were used for 
            correcting the measured spectra yielding intrinsic blazar spectra. }
\end{figure}

\section{Results for Mrk~421 and Mrk~501}
The energy spectra of Mrk~421 and Mrk~501 have been measured with high statistical precision 
by several groups.  We refrain from merging data from different groups, since the 
uncertainties in the spectra are dominated by systematics.   We derive the luminosity peak energies
for the Whipple and HEGRA data individually.  This approach provides an 
estimate  of the systematic uncertainties resulting from the different instrument calibrations
and analyses. 
\begin{figure}[t]
  \begin{center}
    \includegraphics[height=22pc]{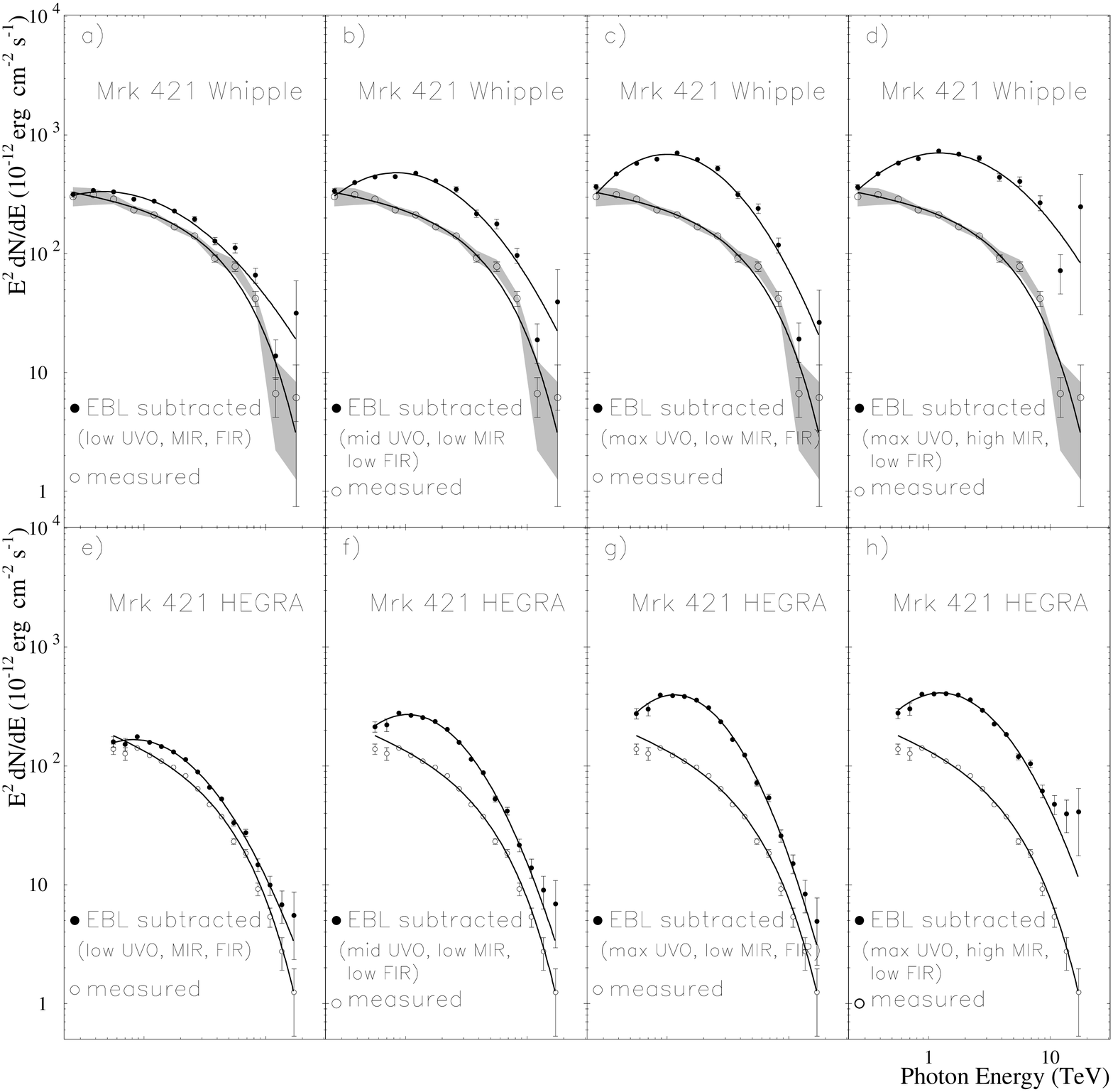}
  \end{center}
  \vspace{-0.5pc}
  \caption{The measured energy spectrum of Mrk~421  in 2000/2001 is shown,
          based on Whipple (Krennrich et al. 2001) and HEGRA data (Aharonian et al. 2002) and the 
          corrected intrinsic spectra are presented for different EBL scenarios.}
\end{figure}
The EBL data is compatible with a range of  $\rm \gamma$-ray absorption opacities and we consider 
various combinations of high and low diffuse photon densities in the optical/ultraviolet (UVO), 
mid-infrared (MIR) and the far-infrared (FIR).  By fitting polynomials through 
the EBL data points (limits or detections) of the various scenarios, we arrived at EBL spectral 
distributions that were used to calculate the optical depths.
This leads to a number of EBL scenarios (see fig. 1) and corrected blazar spectra.

\begin{figure}[t]
  \begin{center}
    \includegraphics[height=22pc]{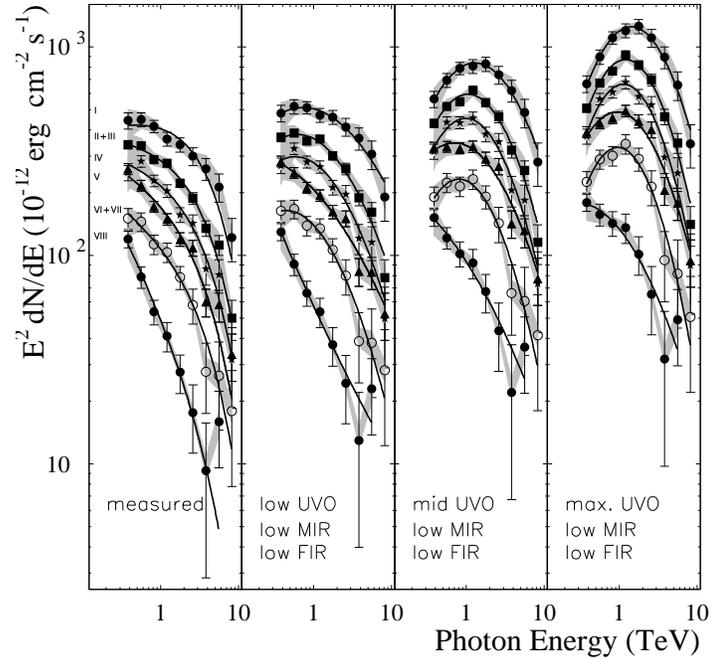}
  \end{center}
  \vspace{-0.5pc}
  \caption{The measured energy spectra of Mrk~421 at different  flux levels averaged for the  2000/2001
            observing season (Krennrich et al. 2002) and the corrected intrinsic spectra are 
             presented for different EBL scenarios are shown.}
\end{figure}

Figure 2~a) shows the energy spectrum of Mrk~421 (empty circles) from data of the Whipple 10~m 
telescope and the spectrum corrected for absorption by an EBL with a low density UVO,  MIR and FIR. 
Figure 2~b) - 2~d) shows the intrinsic spectra for an average UVO, low MIR, low FIR, a high UVO, low MIR, low FIR
and a high UVO, high MIR and low FIR.   Figure 2~e) through 2~h) shows the corresponding spectra 
using the HEGRA data.    
The peak energies have been derived by fitting a parabolic function to the individual spectra.   
The Whipple data yield peak energies between $\rm 468 \pm 52$~GeV and $\rm 1227 \pm 68$~GeV.
The HEGRA data indicate peak energies between $\rm 818 \pm 75$~GeV and $\rm 1252 \pm 81$~GeV.  
For all scenarios but the low UVO, MIR, FIR the HEGRA and Whipple data show compatible results.
Peak energies for  Mrk~501 during strong flares in 1997,  range between $\rm 785 \pm 153$~GeV 
and $\rm 1761 \pm 179$~GeV for the Whipple data and 
between $\rm 1172 \pm 118$~GeV and $\rm 2390 \pm 127$~GeV for the HEGRA data.
In summary, the peak energy for the TeV blazar Mrk~501 is about 30\% higher than for  Mrk~421 when
comparing both in a  flaring state lasting for several months
 
\noindent Figure 3~a) shows the energy spectra of Mrk~421 during a strong flaring state in 2000/2001 with the 
data binned into  different flux levels. Figures 3~b) - 3~d) show the corrected spectra.  
Intrinsic spectra resulting from all three EBL scenarios  indicate a shift in the peak energy 
between the lowest state and the higher flux levels. Further details 
will be given in Dwek \& Krennrich  (2003, in preparation).


\section{References}
\re
Aharonian, F.A. et al., 1999, A\&A, 351, 330
\re
Aharonian, F.A. et al., 2002, A\&A, 384, L23
\re
Gould, R. J., \& Schr\' eder,  G. 1967, Phys. Rev., 155, 1408
\re
Hauser, M.G. \& Dwek, E. 2001, ARA\&A, 39, 249
\re
Kelsall, T. 1998, ApJ, 508, 44
\re
Krennrich, F. et al., 2001, ApJ, 560, L45
\re
Krennrich, F. et al., 2002, ApJ, 575, L9
\re
Partridge, R.B. \& Peebles, P.J.E., 1967, ApJ, 148, 377 
\re
Primack, J.R., 1999,  Astroparticle Physics, 11, 93
\re
Samuelson, F.W. et al. 1998, ApJ, 501, L17
\re
Stecker, F.W., De Jager, O.C., \& Salamon, M.H. 1992, ApJ, 390, L49
\re
\endofpaper
\end{document}